\documentclass[aps, prl, superscriptaddress, reprint, twocolumn, amsmath, amssymb, a4, notitlepage]{revtex4-1}
\usepackage{graphicx}
\usepackage{amsmath,amssymb}
\usepackage{dcolumn}
\usepackage{mathrsfs}
\usepackage{physics}
\usepackage[utf8]{inputenc}

\begin{document}
\title{Fundamental cavity--waveguide interplay in cavity QED}
	\author{Emil V. Denning}
	\email{emvo@fotonik.dtu.dk}

	\affiliation{Department of Photonics Engineering, DTU Fotonik, Technical University of Denmark, Building 343, 2800 Kongens Lyngby, Denmark}

	\author{Jake Iles-Smith}
	\affiliation{Department of Photonics Engineering, DTU Fotonik, Technical University of Denmark, Building 343, 2800 Kongens Lyngby, Denmark}
	\affiliation{School of Physics and Astronomy, The University of Manchester, Oxford Road, Manchester M13 9PL, UK}
	
	\author{Andreas Dyhl Osterkryger}
	\affiliation{Department of Photonics Engineering, DTU Fotonik, Technical University of Denmark, Building 343, 2800 Kongens Lyngby, Denmark}
	
	\author{Niels Gregersen}
	
	\author{Jesper Mork}
	\email{jesm@fotonik.dtu.dk}
	\affiliation{Department of Photonics Engineering, DTU Fotonik, Technical University of Denmark, Building 343, 2800 Kongens Lyngby, Denmark}
	
\begin{abstract}
Interfacing solid-state emitters with photonic structures is a key strategy for developing highly efficient photonic quantum technologies.
Such structures are often organised into two distinct categories: nanocavities and waveguides. 
However, any realistic nanocavity structure simultaneously has characteristics of both a cavity and waveguide, which is particularly pronounced when the cavity is constructed using low-reflectivity mirrors in a waveguide structure with good transverse light confinement.
In this regime, standard cavity quantum optics theory breaks down, as the waveguide character of the underlying dielectric is only weakly suppressed by the cavity mirrors. 
By consistently treating the photonic density of states of the structure, 
 we provide a microscopic description of an emitter including the effects of phonon scattering over the full transition range from waveguide to cavity. 
This generalised theory lets us identify an optimal regime of operation for single-photon sources in optical nanostructures, where cavity and waveguide effects are concurrently exploited. 
\end{abstract}

	\maketitle


Solid-state emitters in photonic nanostructures play an important role in quantum optics and photonic quantum technologies~\cite{obrien2009photonic, kok2007linear,aharonovich2016solid}, both as single-photon sources~\cite{varnava2008good}, and more generally as light-matter interfaces~\cite{arnold2015macroscopic,hu2008giant}.
Such nanostructures can be divided into two generic classes: Nanocavities work by enhancing spontaneous emission into a well-defined cavity mode through the Purcell effect~\cite{Pur46}, while simultaneously suppressing decoherence mechanisms~\cite{he2013demand,ding2016demand,somaschi2016near,kaer2013microscopic,iles2017phonon,grange2017reducing}. 
Waveguides exploit slow-light effects in photonic crystal line defects~\cite{arcari2014near,rao2007single,lecamp2007very} or screening effects in e.g. nanowires~\cite{claudon2010highly,reimer2012bright}, such that spontaneous emission occurs preferentially into the desired channel, thus achieving high efficiencies over a broad frequency range. 
These two classes of structures are treated very differently in standard quantum optics theory. 
Nanocavities are often modelled using a standard Jaynes-Cummings treatment, where the electric field in the cavity is quantised as a single optical mode~\cite{jaynes1963comparison}, while waveguides are modelled as an unstructured reservoir with a continuum of optical modes with little or no spectral variation~\cite{Weisskopf1930}. 
This is a problem in the regime of strongly dissipative cavities, where neither of the two models provide a good physical description. 

\begin{figure}
	\centering
	\includegraphics[width=\columnwidth]{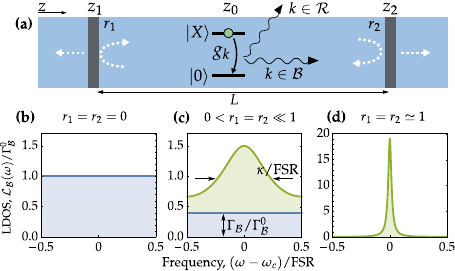}
	\caption{{\bf (a)} Schematic of a two-level emitter in a waveguide structure with two mirrors forming a Fabry-P\'erot cavity. {\bf (b)--(d)} Optical LDOS vs. frequency, scaled with the free spectral range (FSR), at the position of the emitter for mirrors with weak, intermediate and high reflectivity, respectively. }
	\label{fig:intro}
\end{figure}

To illustrate this point, consider a cavity embedded in a waveguide structure defined by mirrors with variable reflectivity (Fig.~\ref{fig:intro}a). 
If the reflectivity of the mirrors is decreased, intuitively one would expect a smooth transition between a strongly localised single-mode cavity, to a standard broadband photonic waveguide, with an intermediate regime at low $Q$ factors, where the optical density of states simultaneously exhibits characteristics of both a waveguide and cavity~\cite{ismail2016fabry,kristensen2013modes,lalanne2018light, barnett1988quantum}. 
However, the Jaynes-Cummings model does not demonstrate this behaviour, failing to describe the properties of emitters in either a waveguide or bulk medium for vanishing $Q$ factors. 

In this paper, we present a quantum optical model that captures the transition between a high-$Q$ cavity and a waveguide, allowing consistent treatment of waveguides, lossy resonators and high quality cavities. 
Our model constitutes a bridge between highly accurate optical simulations of nanostructures~\cite{de2018benchmarking} and microscopic quantum dynamical calculations. This way, the quantum properties of generated light can be calculated, while fully accounting for the electromagnetic properties of the nanostructure.
The generality of this theory enables us to identify an optimal regime of operation for quantum dot single-photon sources, which simultaneously harnesses the high efficiency of a waveguide and the phonon-suppressing spectral structure of a cavity. 




We shall consider a two-level emitter placed in a waveguide with two mirrors forming a Fabry-P\'erot cavity (Fig. \ref{fig:intro}a). We denote the ground and excited states of the emitter by $\ket{g}$ and $\ket{e}$, respectively, separated by the transition frequency $\omega_X$. The electromagnetic field can be described by a set of modes with annihilation and creation operators, $a_k$ and $a_k^\dagger$, frequencies $\omega_k$ and emitter coupling strengths $g_k$. The Hamiltonian governing the entire system takes the form $H=H_\mathrm{E}+H_\mathrm{F}+H_\mathrm{EF}$, where $H_\mathrm{E},\; H_\mathrm{F}$ and $H_\mathrm{EF}$ are the Hamiltonians governing the emitter, electromagnetic field and their interaction, respectively. In the rotating wave approximation, they take the standard forms ($\hbar=1$): $H_\mathrm{E}=\omega_X\dyad{X}$, $H_\mathrm{F}=\sum_k \omega_k a_k^\dagger a_k$, and $H_\mathrm{EF}=\sum_k g_k a_k^\dagger \sigma + \mathrm{H.c.}$, with $\sigma=\dyad{g}{e}$. 
The $k$ indices labelling the optical modes can be divided into two sets: the first set, $\mathcal{B}$, contains all the modes with a certain transverse field profile of interest, for example that of the fundamental waveguide mode; the second set, $\mathcal{R}$, accounts for non-guided radiation modes, and guided modes with different transverse field profile if the structure is not single-moded. 
Each mode set has an associated local density of states (LDOS), 
\begin{align}
\mathcal{L}_\mathcal{S}(\omega) = \pi\sum_{k\in\mathcal{S}} \abs*{g_k}^2\delta(\omega-\omega_k), \;\; \mathcal{S}=\mathcal{B},\mathcal{R}.
\end{align}
In the absence of mirrors in the waveguide, both densities can be considered constant over a large frequency range, $\mathcal{L}_\mathcal{S}\simeq \Gamma_\mathcal{S}^0$, where $\Gamma_\mathcal{S}^0$ is the spontaneous emission rate of the emitter into the mode set $\mathcal{S}$ (cf. Fig.~\ref{fig:intro}b). However, when mirrors with amplitude reflectivities $r_1,\;r_2$ are added to the structure, $\mathcal{L}_\mathcal{B}$ becomes~\cite{gregersen2016broadband}
\begin{align}
\label{eq:LDOS-exact}
\mathcal{L}_\mathcal{B}(\omega) = \Gamma_\mathcal{B}^0  \Re\qty{\frac{[1+\tilde{r}_1(\omega)][1+\tilde{r}_2(\omega)]}{1-\tilde{r}_1(\omega)\tilde{r}_2(\omega)}},
\end{align}
where $\Gamma_\mathcal{B}^0$ is the emission rate into the waveguide in the absence of mirrors, which is highly dependent on the local field strength and position of the emitter.
The phase accumulated during propagation in the cavity is given by the effective complex reflectivity coefficient $\tilde{r}_j(\omega)=r_j e^{i[\phi_0^j + L\beta(\omega)]}$, where $\phi_0^j$ is a mirror reflection phase, $L$ is the cavity length, and we assume a dispersion-less propagation factor, $\beta(\omega)=n_\mathrm{eff}\omega/c$, where $n_\mathrm{eff}$ is the effective refractive index of the waveguide mode. For simplicity, we have assumed that the emitter is placed in the middle of the cavity. Further, since the mirror reflection phase only amounts to a shift in resonance and the position of the field antinodes in the cavity, they may be safely neglected.

Importantly, for intermediate reflectivities, the LDOS features a Lorentzian lineshape offset by a constant background~\cite{gregersen2016broadband}, as shown in Fig. \ref{fig:intro}c, where \eqref{eq:LDOS-exact} is plotted for $r_1=r_2=0.2$.  The background contribution to the LDOS, $\Gamma_\mathcal{B}$, stems from the waveguide nature of the dielectric structure, while the Lorentzian peak is a signature of the cavity quasi-mode, which becomes the dominant contribution to the LDOS as $r\simeq 1$ (cf. Fig~\ref{fig:intro}d). To separate these two contributions from each other, we approximate the LDOS, \eqref{eq:LDOS-exact}, as $\mathcal{L}_\mathcal{B}(\omega)\simeq \bar{\mathcal{L}}_\mathcal{B}(\omega)=\Gamma_\mathcal{B} + \mathcal{L}_c\frac{\tilde{\kappa}}{\tilde{\kappa}^2+\tilde{\omega}^2}$,
where we have introduced the dimensionless frequency $\tilde{\omega}=Ln_\mathrm{eff}(\omega-\omega_c)/c$ and linewidth, $\tilde{\kappa}=\kappa (L n_\mathrm{eff}/c)$. The LDOS weights $\Gamma_\mathcal{B}$ and $\mathcal{L}_c$ determine the contributions from background waveguide modes and the cavity, respectively. In the Supplemental Information (SI), we show how the mirror reflectivities $r_1$ and $r_2$, and the bare waveguide emission rate $\Gamma_\mathcal{B}^0$ uniquely determine all three parameters $\Gamma_\mathcal{B}$, $\mathcal{L}_c,$ and $\tilde{\kappa}$. Furthermore, $\mathcal{L}_c$, $L$, and $n_\mathrm{eff}$ determine the emitter--cavity coupling strength, $g=\sqrt{c\mathcal{L}_c /(4Ln_\mathrm{eff})}$. 

\begin{figure}
	\centering
	\includegraphics[width=\columnwidth]{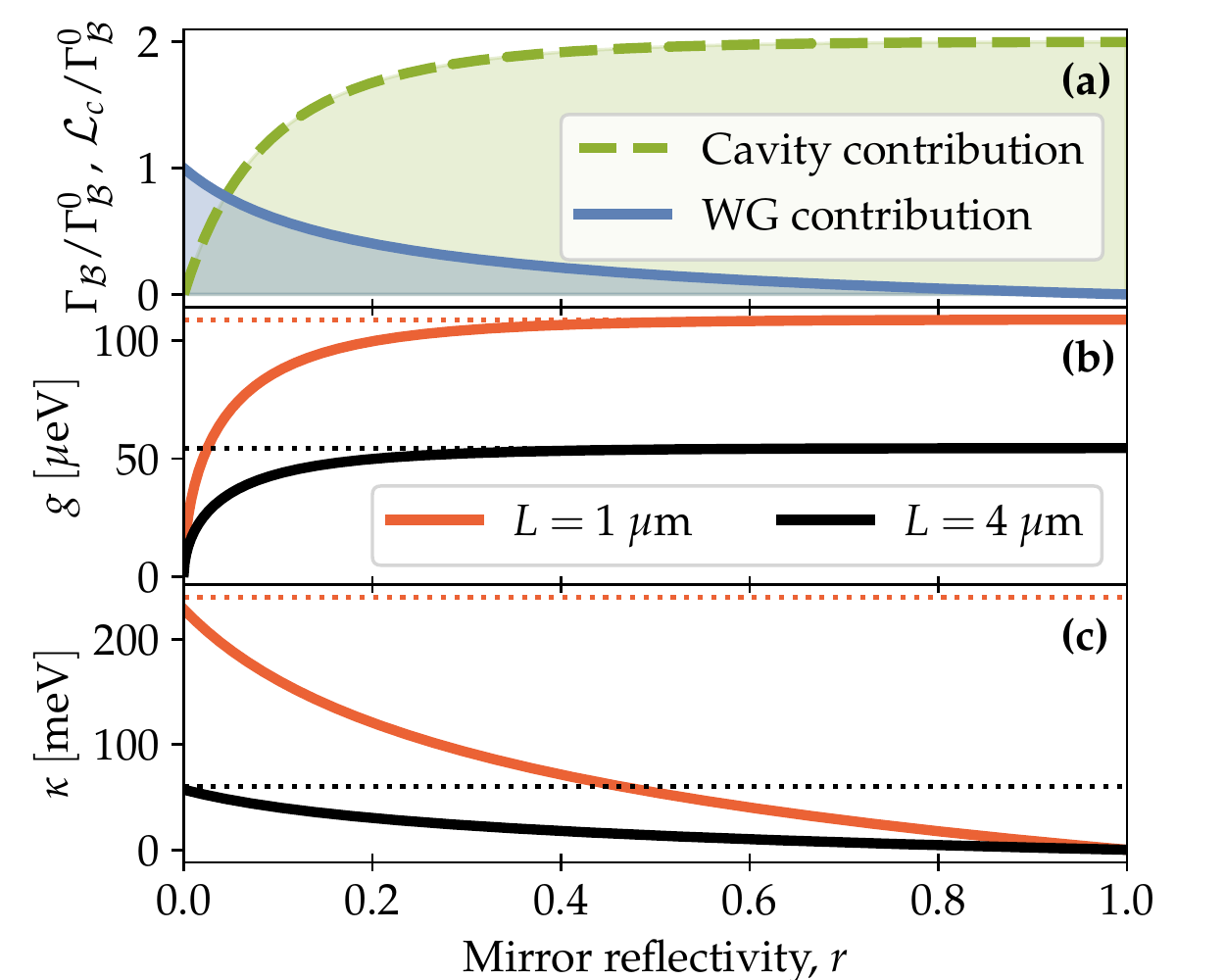}
	\caption{{\bf (a)} Contributions to LDOS from waveguide, $\Gamma_\mathcal{B}$ (solid) and  cavity, $\mathcal{L}_c$ (dashed) vs. mirror reflectivity, $r_1=r_2\equiv r$. {\bf (b)}, {\bf (c)} Dependence of emitter--cavity coupling rate, $g$, and cavity decay rate, $\kappa$, on mirror reflectivity for cavities with $L=1\;\mu \mathrm{m}$ (red) and $L=4\;\mu \mathrm{m}$ (black). Other parameters: $\Gamma_\mathcal{B}^0=0.3\;\mu\mathrm{eV},\; n_\mathrm{eff}=2.5$. The limiting values  $g_\mathrm{max}$ and $\kappa_\mathrm{max}$ are indicated in (b) and (c) with dotted lines.}
	\label{fig:g-kappa}
\end{figure}

In Fig.~\ref{fig:g-kappa}a, the contribution to the LDOS from waveguide background modes and the cavity quasimode is shown as functions of the mirror reflectivity for a symmetric cavity ($r_1=r_2\equiv r$). These depend solely on the mirror reflectivity and show clearly how the system is gradually transformed from a waveguide for $r=0$, to a cavity with full suppression of the waveguide background as $r\rightarrow 1$. Similarly, the variation of the emitter--cavity coupling strength is shown in Fig.~\ref{fig:g-kappa}b. While $g$ is normally assumed to depend only on the cavity mode volume, $V$, we see here that $g\rightarrow 0$ as $r\rightarrow 0$, since the cavity does not contribute to the LDOS in this limit. As the reflectivity increases, $g$ approaches the value $g_\mathrm{max}=\sqrt{\Gamma_\mathcal{B}^0 c/(2L n_\mathrm{eff})}$ (dotted lines in Fig.~\ref{fig:g-kappa}b). This is consistent with the conventional scaling of $g$ as $\sim 1/\sqrt{V}$, noting that $\Gamma_\mathcal{B}^0\sim 1/A$ with $A$ the transverse mode area.
Fig.~\ref{fig:g-kappa}c shows how the cavity linewidth tends to zero as the mirror reflectivity is increased. Importantly, the linewidth does not diverge as $r\rightarrow 0$, but rather approximately converges to the value $\kappa_\mathrm{max}=2c/(n_\mathrm{eff} L)$ (dotted lines in Fig.~\ref{fig:g-kappa}c), which is the inverse of the time it takes for light to propagate from the middle of the cavity to one of the mirrors~\cite{rao2007single}. 

From these parameters, we are able to derive a quantum optical master equation to describe the dynamical and optical properties of the emitter (see SI for details),
\begin{align}
\label{eq:master-eq}
\begin{split}
\dot{\rho}(t) &= -i[g(\hat{a}^\dagger\sigma+\hat{a}\sigma^\dagger),\rho(t)]
+ (\Gamma_\mathcal{B}+\Gamma_\mathcal{R}) \mathcal{D}[\sigma]+ \kappa\mathcal{D}[\hat{a}].
\end{split}
\end{align}
where $\hat{a}$ ($\hat{a}^\dagger$) is the annihilation (creation) operator for the cavity mode, and $\mathcal{D}[x]=x\rho(t)x^\dagger - \frac{1}{2}\{x^\dagger x,\rho(t) \}$ is the Lindblad dissipator. 
In the vanishing mirror limit, $r= 0$, we have $g=0, \; \Gamma_\mathcal{B}=\Gamma_\mathcal{B}^0$, and the master equation reduces to the usual waveguide case. 
Conversely, in the high reflectivity limit, $r\rightarrow 1$, the waveguide contribution to the LDOS vanishes, $\Gamma_\mathcal{B}\rightarrow 0$ and the master equation describes an emitter coupled to a cavity quasimode and a radiation bath.

If the waveguide structure is single-moded, the emission rate $\Gamma_\mathcal{R}$ only accounts for emission into radiation modes out of the waveguide structure, and it can be taken independent of the mirror reflectivity, $\Gamma_\mathcal{R}\simeq\Gamma_\mathcal{R}^0$. 
However, if the structure is multi-moded, $\Gamma_\mathcal{R}$ also accounts for emission into waveguide modes with different transverse field distribution than $\mathcal{B}$. The cavity mirrors also modulate the LDOS for these modes, such that the total LDOS of the radiation reservoir, $\mathcal{R}$, becomes
\begin{align}
\mathcal{L}_\mathcal{R}(\omega) = \Gamma_\mathrm{RM} + \sum_m \Gamma_m^0 \Re\qty{\frac{[1+\tilde{r}^m_1(\omega)][1+\tilde{r}^m_2(\omega)]}{1-\tilde{r}^m_1(\omega)\tilde{r}^m_2(\omega)}},
\end{align}
where $\Gamma_\mathrm{RM}$ is the emission rate into radiation modes and the sum runs over all other mode families in the waveguide, except for the $\mathcal{B}$ mode of interest. In the absence of cavity mirrors, we have the spontaneous emission rate, $\Gamma_m^0$, and complex reflectivity, $\tilde{r}_j^m(\omega)=r_j^m e^{i[\phi_j + L\beta_m(\omega)]}$, associated to the $m^{\text{th}}$ mode, where $\beta_m$ is the corresponding propagation constant. Presuming that the emitter is only resonant with the mode of interest, we can assume weak coupling to the remaining modes, such that the emitter decay into $\mathcal{R}$ is simply described by the spontaneous emission rate $\Gamma_\mathcal{R}=\mathcal{L}_\mathcal{R}(\omega_X)$.

We now apply our formalism to the case of a quantum dot (QD) single-photon source in a dielectric waveguide structure with mirrors, taking scattering with longitudinal acoustic phonons into account. We take one cavity mirror to be perfectly reflecting and the other to have a finite reflectivity, $r_1=1,\; r_2\equiv r$. Due to interference effects, the presence of a perfectly reflecting mirror modulates the LDOS by a sinusoidal variation with a period of the free spectral range, $Ln_\mathrm{eff}/c$, even in the limit $r=0$. If the frequency range of interest is appreciably smaller than this range, as is often the case for a QD in a nanocavity, we find that the effect of the perfect bottom mirror can be implemented by using the renormalised rates, $\Gamma_\mathcal{B}^{0*}=2\Gamma_\mathcal{B}^0$ and $\kappa^*=\kappa/2$, where $\Gamma_\mathcal{B}^0$ and $\kappa$ are calculated assuming a symmetric cavity as in Fig.~\ref{fig:g-kappa}. This means that in the limit of a vanishing front mirror reflectivity, the renormalised $\beta$ factor in the presence of the back mirror is $\beta^*=2\Gamma_\mathcal{B}^0/(2\Gamma_\mathcal{B}^0+\Gamma_\mathcal{R})=2\beta/(\beta+1)$. Here, $\beta=\Gamma_\mathcal{B}^0/(\Gamma_\mathcal{B}^0+\Gamma_\mathcal{R})$ is the waveguide $\beta$ factor in the absence of both mirrors, not to be confused with the wave propagation constant.
Furthermore, we assume that the underlying waveguide structure is single-moded such that $\Gamma_\mathcal{R}=\Gamma_{\mathrm{RM}}$ can be considered constant. 

The total Hamiltonian of the system is given by $H=H_\mathrm{E}+H_\mathrm{F}+H_\mathrm{EF}+H_\mathrm{P}+H_\mathrm{EP}$, with $H_\mathrm{P}$ and $H_\mathrm{EP}$ the free phonon and emitter--phonon Hamiltonians, given by~\cite{mahan2013many,ramsay2010phonon}
\begin{align}
H_\mathrm{P}= \sum_\mathbf{q} \nu_\mathbf{q} b_\mathbf{q}^\dagger b_\mathbf{q},\; H_\mathrm{EP}= \dyad{e}\sum_\mathbf{q} M_\mathbf{q}(b_\mathbf{q}+b_\mathbf{q}^\dagger),
\end{align}
where $b_\mathbf{q}$ ($b_\mathbf{q}^\dagger$) is the annihilation (creation) operator for the phonon mode with wavevector $\mathbf{q}$, with associated frequency $\nu_\mathbf{q}$, and exciton coupling strength $M_\mathbf{q}$. 
The phononic spectral density is given by $\mathcal{J}(\nu)=\sum_\mathbf{q} M_\mathbf{q}^2\delta(\nu-\nu_\mathbf{q})=\alpha \nu^3 \exp[-\nu^2/\nu_c^2]$, where $\alpha$ is the exciton--phonon coupling parameter and $\nu_c$ the cutoff frequency~\cite{nazir2016modelling}. 
To calculate the dynamics and account for non-Markovian phonon relaxation, we make use of the polaron theory~\cite{mccutcheon2010quantum,mccutcheon2011general,roy2011influence,wilson2002quantum,nazir2016modelling, iles2017phonon}. 
This is done by first applying the unitary transformation $\mathcal{T}=\exp(\dyad{e}S)$ 
to the Hamiltonian, $H\rightarrow \hat{H}=\mathcal{T}H\mathcal{T}^\dagger$, where $S = \sum_\mathbf{q}\nu_\mathbf{q}^{-1}M_\mathbf{q}(b_\mathbf{q}^\dagger - b_\mathbf{q})$. 
 In this frame, an equation of motion for the reduced state of the QD that is non-perturbative in the electron-phonon coupling strength may be derived (see SI for details).
For completeness, we also include a pure dephasing process~\cite{grange2017reducing} with rate $\gamma$, which accounts for dephasing from charge noise~\cite{Thoma}, spin noise and virtual electron--phonon scattering~\cite{muljarov2004dephasing,reigue2017probing}. 

\begin{figure}
	\centering
	\includegraphics[width=0.9\columnwidth]{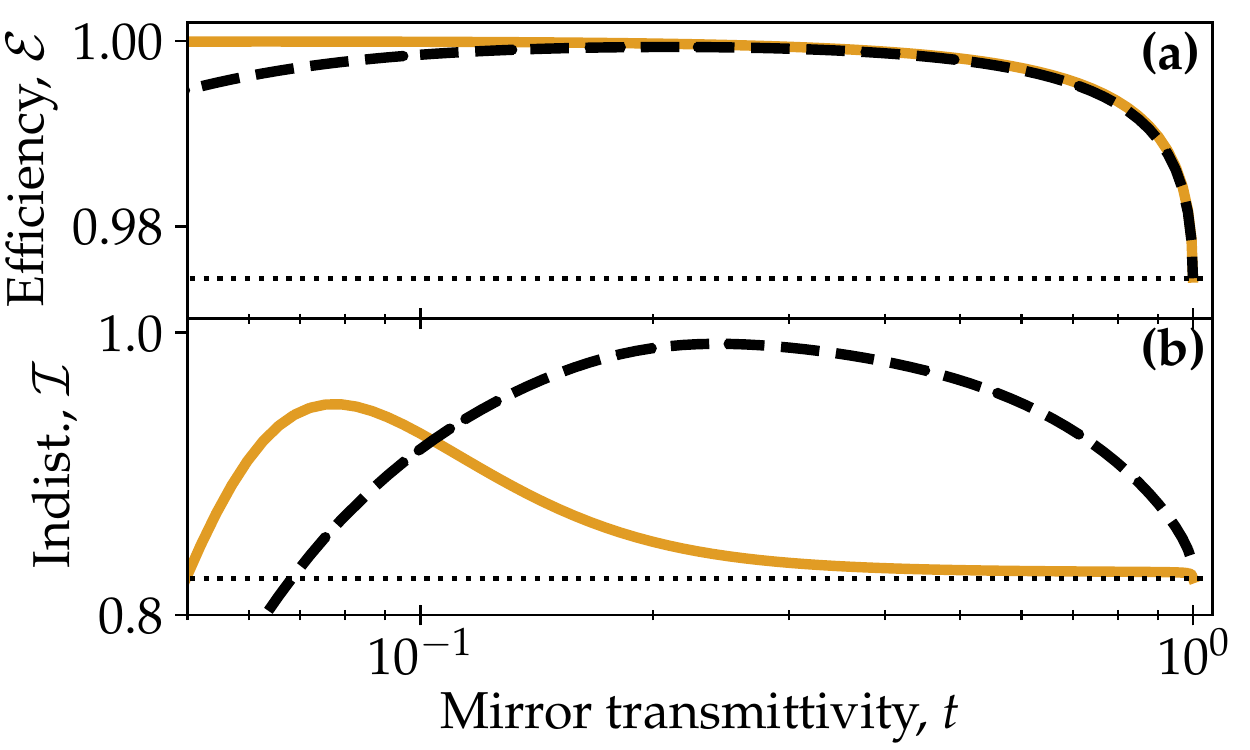}
	\caption{ {\bf (a)} Efficiency and {\bf (b)} indistinguishability of single photon source vs. transmittivity of top mirror for cavities with lengths $L=\frac{1}{2}\lambda_X/n_\mathrm{eff}$ (solid) and $L=15\lambda_X/n_\mathrm{eff}$ (dashed). The thin dotted line indicates $\beta^*=0.974$ in (a) and $B^4=0.826$ in (b). The spontaneous emission rate into the mirrorless waveguide is $\Gamma_\mathcal{B}^0=1.1\; \mu\mathrm{eV}$.}
	\label{fig:eff-ind}
\end{figure}

The indistinguishability of photons emitted from the QD  into detected modes~\cite{kiraz2004quantum,iles2017phonon} can be calculated using tools from optical theory~\cite{novotny2012principles,wubs2004multiple}, 
as discussed in the SI. This leads to
\begin{align}
\label{eq:indist}
\mathcal{I}=[2\mathcal{P}_\mathcal{B}/\Gamma_\mathcal{B}^0]^{-2}\int_{-\infty}^\infty \dd{\omega}\dd{\omega'} \abs{\mathcal{G}^*(\omega)\mathcal{G}(\omega')S_0(\omega,\omega')}^2,
\end{align}
where $\mathcal{G}(\omega)=[1+\tilde{r}_1(\omega)]t_2[1-\tilde{r_1}(\omega)\tilde{r}_2(\omega)]^{-1}$ accounts for the cavity filtering, 
 $\mathcal{P}_\mathcal{B}=(\Gamma_\mathcal{B}^0/2)\int_{-\infty}^\infty\dd{\omega} \abs*{\mathcal{G}(\omega)}^2 S_0(\omega,\omega)$ is the total power in the $\mathcal{B}$ modes, and $S_0(\omega,\omega')=\int_{-\infty}^\infty\dd{t}\dd{t'}e^{i(\omega t-\omega't')}\ev*{\sigma^\dagger(t)\sigma(t')}$ is the two-colour dipole spectrum. 
The two-colour spectrum may be separated into two contributions $S_0=S_\mathrm{ZPL}+S_\mathrm{PSB}$, where $S_\mathrm{ZPL}$
describes emission into a sharp zero-phonon line (ZPL) from direct exciton relaxation, and $S_\mathrm{PSB}$ corresponds to a broad phonon sideband (PSB) where a phonon and a photon are simultaneously emitted. Centrally, the Franck-Condon factor, $B^2$, is the fraction of photons emitted into the ZPL, if the electromagnetic LDOS is frequency independent~\cite{iles2017phonon}. Here, $B=\mathrm{Tr} [B_+ e^{-H_\mathrm{P}/(k_B T)}]/\mathrm{Tr}[e^{-H_\mathrm{P}/(k_BT)}]$, where $T$ is the temperature and $k_B$ is the Boltzmann constant.
The efficiency is defined as the ratio of energy emitted into the desired waveguide mode and the total emitted energy. It is calculated as $\mathcal{E}=\mathcal{P}_\mathcal{B}/(\mathcal{P}_\mathcal{B}+\mathcal{P}_\mathcal{R})$, where $\mathcal{P}_\mathcal{R}=\Gamma_\mathcal{R}\int_{-\infty}^\infty\dd{\omega} S_0(\omega,\omega)$.
In Fig.~\ref{fig:eff-ind}, the efficiency and indistinguishability are plotted versus transmittivity of the finitely reflecting cavity mirror for identical waveguide structures with cavity lengths $L=\frac{1}{2}\lambda_X/n_\mathrm{eff}$ (solid) and $L=15\lambda_X/n_\mathrm{eff}$ (dashed), where $\lambda_X$ is the QD transition wavelength, taken as $950\mathrm{\; nm}$. For clarity, $\gamma$ has been set to 0. 

In the weak emitter--cavity coupling regime, the cavity resonance is broad compared to the ZPL, but can still vary appreciably over the PSB, meaning phonon-assisted QD relaxation is suppressed and the ZPL is Purcell enhanced. In this parameter regime, we may generalise the results of Ref.~\cite{iles2017phonon} to obtain analytical expressions taking into account the waveguide--cavity interplay,
\begin{align}
\label{eq:I-E-analytical}
\mathcal{I}
&=\frac{\Gamma_\mathrm{tot}}{\Gamma_\mathrm{tot}+ 2\gamma_\mathrm{tot}}\qty[\frac{(\Gamma_\mathcal{B}+\Gamma_\mathrm{cav})B^2}{(\Gamma_\mathcal{B}+\Gamma_\mathrm{cav})B^2+2\Gamma_\mathcal{B}^0F(1-B^2)}]^2\\
\mathcal{E}
\label{eq:I-E-analytical-E} &=\frac{(\Gamma_\mathrm{cav}+\Gamma_\mathcal{B})B^2+2\Gamma_\mathcal{B}^0F(1-B^2)}{(\Gamma_\mathrm{cav}+\Gamma_\mathcal{B})B^2+2\Gamma_\mathcal{B}^0F(1-B^2) +\Gamma_\mathcal{R}},
\end{align}
where $F=[\int\dd{\omega}S_\mathrm{PSB}(\omega,\omega)]^{-1}\frac{1}{4}\int\dd{\omega}\abs*{\mathcal{G}(\omega)}\!^2 S_\mathrm{PSB}(\omega,\omega)$ 
is the fraction of the PSB not removed by filtering imposed by the electromagnetic LDOS, $\gamma_\mathrm{tot}=\gamma+2\pi (gB/\kappa)^2\mathcal{J}(2gB)\coth(gB/(k_B T))$ is a phonon-enhanced pure dephasing rate and $\Gamma_\mathrm{tot}=\Gamma_\mathrm{cav}+\Gamma_\mathcal{B}+\Gamma_\mathcal{R}$. In the limit $t=1$, \eqref{eq:I-E-analytical-E} reduces to $\mathcal{E}=\Gamma_\mathcal{B}^{0*}/(\Gamma_\mathcal{B}^{0*}+\Gamma_\mathcal{R})$, meaning the efficiency converges towards $\beta^*$, as the front mirror is gradually removed (black dotted line in Fig.~\ref{fig:eff-ind}a). If the waveguide mode contribution to the LDOS were ignored, the efficiency would approach zero as $r\rightarrow 0$~\cite{iles2017phonon}, which is only a valid approximation when the underlying waveguide structure has a vanishing $\beta$ factor. In the absence of pure dephasing, $\gamma=0$, \eqref{eq:I-E-analytical} gives $\mathcal{I}\rightarrow B^4$ (black dotted line in Fig.~\ref{fig:eff-ind}b) as $t\rightarrow 1$, consistent with Ref.~\cite{iles2017phonon}.

Contrarily, in the strong emitter--cavity coupling regime, $g>\kappa$, the cavity and exciton hybridise and form a polariton pair. In this case, the phonons drive incoherent transitions between the two polaritons, leading to a decreased photon indistinguishability. To resolve this effect, quantisation of the cavity becomes crucial, and a semi-classical weak light-matter coupling theory is insufficient. As seen in Fig.~\ref{fig:eff-ind}b, increasing the cavity length leads to a smaller $g$, which allows further narrowing of the cavity line and thus suppression of the phonon sideband without entering the strong coupling regime, due to a larger cavity mode volume. Since the underlying waveguide structure has a high $\beta$ factor, the efficiency does not suffer noticeably from this. The efficiency starts to decrease when $\kappa$ becomes small enough that the photon escapes the cavity by scattering to radiation modes via the QD rather than dissipating through the mirror. Increasing the cavity length will continue to improve the coherence of emitted photons until the cavity free spectral range becomes comparable to the width of the PSB, which will then become Purcell enhanced. For typical QDs, the PSB extends over a few meV, and the cavity would need a length of $50-100\,\mu\mathrm{m}$ for this effect to set in.

In conclusion, we have characterised the important role of weakly suppressed waveguide modes in nanocavities. As a demonstration of this, we have shown that long nanocavities based on high $\beta$ factor waveguides constitute a promising new route to high-performance single photon sources. 

\begin{acknowledgements}
E.V.D, J.I.S., and J.M. acknowledge funding from the Danish Council for Independent Research (Grant No. DFF-4181-00416) and from Villum Fonden (Grant 8692). J.I.S is supported by the Engineering and Physical Sciences Research Council, Grant No. EP/N008154/1. 
N.G and A.D.O. acknowledge support from the Danish National Research Council for Technology and Production (LOQIT Sapere Aude grant DFF \# 4005-00370).
\end{acknowledgements}

\end{document}